\documentclass[12pt,reqno]{amsart}
\usepackage{amssymb,amsfonts,latexsym,fullpage,graphicx}

\theoremstyle{plain}

\newtheorem{prop}{\bf{Proposition}}

\theoremstyle{definition}

\newcommand{\roundbrackets}[1]{\left(#1\right)}
\newcommand{\squarebrackets}[1]{\left[#1\right]}

\newcommand{\anglebrackets}[1]{\left\langle#1\right\rangle}

\newcommand{\ft}[1]{{#1}}

\newcommand{\sinc}[1]{\text{sinc}\left(#1\right)}
\renewcommand{\sin}[1]{\text{sin}\left(#1\right)}
\renewcommand{\cos}[1]{\text{cos}\left(#1\right)}

\newcommand{\conj}[1]{#1^{*}}
\newcommand{\abs}[1]{\left|#1\right|}
\renewcommand{\Re}[1]{\text{Re}\left(#1\right)}
\renewcommand{\Im}[1]{\text{Im}\left(#1\right)}

\newcommand{\normtwo}[1]{\left\|#1\right\|}

\newcommand{\ip}[2]{\anglebrackets{#1 \vert #2}}

\newcommand{\ket}[1]{ \left \vert #1 \right \rangle }

\newcommand{\pmax}{p_{\text{max}}}
\newcommand{\lmin}{\lambda_{\text{min}}}
\newcommand{\pmaxint}{\int_{-\pmax}^{\pmax}}
\newcommand{\pmaxinterval}{\squarebrackets{-\pmax,\pmax}}

\newcommand{\slitinterval}{\squarebrackets{-\frac{L}{2},\frac{L}{2}}}
\newcommand{\infint}{\int_{-\infty}^{\infty}}
\newcommand{\kpsi}{\ket{\psi}}

\newcommand{\kPsi}{\ket{\Psi}}
\newcommand{\kPsiN}{\ket{\Psi_{N}}}
\newcommand{\kPhi}{\ket{\Phi}}

\newcommand{\expectation}[1]{\overline{#1}}
\newcommand{\variance}[1]{ \Delta #1 }

\newcommand{\figwid}{150pt}

\newcommand{\p}{{\bf p}}

\begin{document}

$$$$

\centerline{\large \bf Analysis of Superoscillatory Wave Functions \rm}
$$$$
\centerline{\large Matt~S.~Calder, Achim~Kempf}
$$$$
\centerline{ Department of Applied Mathematics, University of Waterloo}

\centerline{ 200 University Avenue West, Ontario N2L 3G1, Canada }

$$$$$$$$
{\small {\bf Abstract.} ~~\quad Surprisingly, differentiable functions are
able to oscillate arbitrarily faster than their highest Fourier component
would suggest. The phenomenon is called \it superoscillation. \rm
Recently, a practical method for calculating superoscillatory functions
was presented and it was shown that superoscillatory quantum mechanical
wave functions should exhibit a number of counter-intuitive physical
effects. Following up on this work, we here present more general methods
which allow the calculation of superoscillatory wave functions with
custom-designed physical properties. We give concrete examples and we
prove results about the limits to superoscillatory behavior. We also give
a simple and intuitive new explanation for the exponential computational
cost of superoscillations.  \rm}

%%%%%%%%%%%%%%%%%%%%%%%%%%%%%%%%%%%%%%%%%%%%%%%%%%%%%%%%%%%%%%%%

\newpage
%\tableofcontents
%\newpage
\section{\bf{Introduction}}

It used to be believed that a function could not oscillate much faster
than its highest Fourier component. Aharonov, Berry and others showed that
this is not the case by giving explicit counter-examples which they named
superoscillatory functions, see, e.g.,
\cite{Aharonov1990}-\cite{Berry1995}. In fact, there are functions which
on arbitrarily long stretches oscillate arbitrarily faster than their
highest frequency Fourier component, see \cite{ak-beet}. In other words,
the presence of localized fast oscillations in a continuous function need
not be visible at all in the function's global Fourier transform. In a
function's global Fourier transform, contributions from regions of fast
oscillations can be cancelled perfectly by contributions from regions
where the wave function is oscillating slowly.

In the context of quantum theory, wave functions that
superoscillate are able to cause a number of counter-intuitive
effects. Some of these may be of conceptual significance in
quantum gravity, see \cite{Kempf2003,Kempf2003b}. But effects of
superoscillations also enter in the low energy realm of
nonrelativistic quantum mechanics. Among such potentially
observable low-energy effects is the counter-intuitive phenomenon
that particles with superoscillatory wave functions can be made to
accelerate when passing through a neutral slit:

Consider a particle which possesses a bounded momentum range, i.e., its
momentum wave function vanishes for momenta that are larger than some
$\pmax$. As will be explained below, we can arrange that in a certain
region in space the particle's wave function superoscillates, i.e. that it
oscillates with a much shorter wavelength than $h/\pmax$. Now let the wave
function be incident onto a screen with a single slit in such a way that
it is the superoscillatory part of the wave function which passes through
the slit. Upon emerging from the slit the particle's wave function will
then oscillate rapidly where the slit is and will be zero elsewhere. The
very short wavelengths of the emerging wave function will be visible in
its global Fourier transform. This is because the contributions to the
global Fourier transform which come from the fast oscillations in the slit
interval are no longer cancelled by contributions from outside the slit
interval. Therefore, the particle will have gained momentum merely by
passing the slit. The momentum gain is determined by the shortness of the
wave-length of the superoscillations and, as explained below, there is no
limit, in principle, to how short that wavelength can be made.

In order to facilitate the design of experiments that can realize the
effects of superoscillatory wave functions it is desirable to possess
methods for explicitly calculating superoscillatory wave functions with
predetermined properties. In particular, one may wish to calculate those
low-momentum but superoscillatory wave functions which after passing
through the slit yield wave functions with a predetermined arbitrarily
large momentum and a momentum uncertainty that is as small as is allowed
by the uncertainty relation. Our aim here is to develop methods that allow
us to solve this and other problems.

\begin{figure}[h]
    \centerline{\includegraphics[width=\figwid,angle=-90]{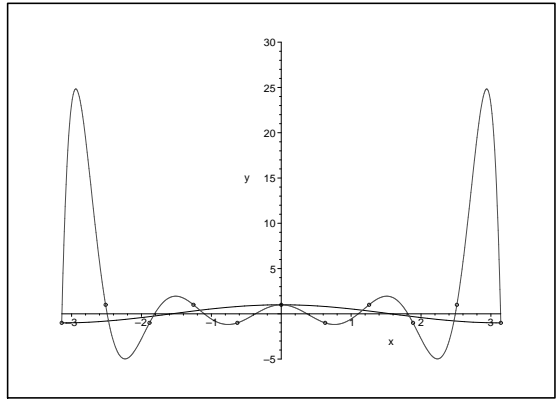}}
    \caption{\label{F:example1}Example of a superoscillation created
    by requiring the wave function to pass through certain points.
    A cosine function of the minimum wavelength/maximum frequency is
    shown for comparison.}
\end{figure}

\begin{figure}[h]
    \centerline{\includegraphics[width=\figwid,angle=-90]{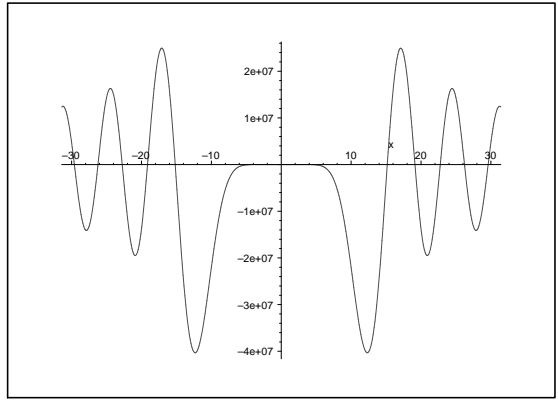}}
    \caption{\label{F:example2}A zoomed-out figure~\ref{F:example1}.
    Notice that, as is typical, the amplitudes in the
    superoscillating region are far smaller (here even unnoticable)
    than those on either side.}
\end{figure}

Our starting point will be the method for calculating superoscillatory
wave functions which was developed in \cite{Kempf2003} using results of
\cite{Levi1965} and \cite{Slepian1978}. This method allows the
construction of wave functions of arbitrarily low fixed frequency content
that pass through an arbitrary finite number of pre-specified points.
Figures~\ref{F:example1} and \ref{F:example2} show an example.

Our aim is to develop more general methods for designing superoscillatory
functions with generic pre-specified properties. We will also ask what the
in-principle limits are for the construction of superoscillatory wave
functions.

\section{\bf{Self-acceleration through single slit}}
In order to motivate and formalize the mathematical problem that we will
address, let us consider the illustrative example of particles that
self-accelerate when passing through a slit.
\subsection{Notation}
We will denote a wave function $\psi$ and its Fourier transform by the
same letter, since they correspond to the same vector in the Hilbert space
of states.
\begin{equation}
    \psi(p) = \frac{1}{\sqrt{2\pi\hbar}} \infint \psi(x) ~e^{-\frac{ipx}{\hbar}} \ dx
\end{equation}
In the text, whenever necessary, we will write $\psi(x)$ or
$\psi(p)$ to indicate if the position or the momentum wave
function is meant. We will often consider particles whose momentum
is bounded by a finite value $\pmax$:
\begin{equation}
    \psi(x) = \frac{1}{\sqrt{2\pi\hbar}} \pmaxint \psi(p)~ e^{\frac{ipx}{\hbar}} \ dx
\end{equation}
Borrowing terminology from communication engineering and sampling theory,
see e.g. \cite{Ferreira2001}, we will speak of such a function $\psi(x)$
as having bandwidth $\pmax$, as being band-limited, or in this case as
being momentum-limited. It will be convenient to define the sinc function
as:
\begin{equation}
    \sinc{x} :=
    \begin{cases}
        \frac{\sin{x}}{x}, ~~\mbox{if}~~x \neq 0 \\
        1, ~~\mbox{if}~~x = 0
    \end{cases}
\end{equation}
Notice that definitions of the sinc function elsewhere may include a
factor of $\pi$.

\subsection{Gedanken experiment}
Let us consider a particle in two dimensions which travels along
the $x_1$-direction towards a screen which is parallel to the
$x_2$-direction. Assume the particle passes through a slit with
$x_2$-coordinate interval $[-\frac{L}{2},\frac{L}{2}]$ in the
screen. Henceforth, we will assume that the incident particle's
momentum parallel to the screen, $\p_2$, has a finite bound
$p_{2_{\text{max}}}$:
\begin{equation}
\ft{\psi}(p_1,p_2,t)  =  0
~~~~~~~~~~\mbox{~~~~~~~if~~~~~~~}~~~~~~~p_2 \notin
[-p_{2_{\text{max}}},p_{2_{\text{max}}}]
\end{equation}
Our aim is to compare the particle's momentum parallel to the
screen before and after the particle passes the slit. For
simplicity, we will suppress the variables $x_1,p_1$ and $t$. From
now on, $x_2$ is renamed $x$ and $p_2$ is renamed $p$. We denote
the incident wave function just before passing through the slit by
$\psi(x)$ and we denote the wave function which emerges from the
slit by $\Psi(x)$. The state $\vert\Psi\rangle$ is of course given
by projection and renormalization
\begin{equation}
    \kPsi := \frac{P_{s}\kpsi}{\normtwo{~P_{s}\kpsi~}}
\end{equation}
where $P_{s}$ projects onto the slit:
\begin{equation}
    P_{s}\psi(x) :=
    \begin{cases}
        \psi(x), ~~\mbox{if}~~\abs{x} \leq \frac{L}{2} \\
        0, ~~\mbox{otherwise}
    \end{cases}
\end{equation}
Similarly, we define $P_{b}$ as the projector onto a finite momentum
interval:
\begin{equation}
    P_{b}\ft{\psi}(p) :=
    \begin{cases}
        \ft{\psi}(p), ~~\mbox{if}~~\abs{p} \leq \pmax \\
        0, ~~\mbox{otherwise}
    \end{cases}
\end{equation}
While the incident wave function is momentum limited,
$P_b\vert\psi\rangle=\vert\psi\rangle$, the emerging wave function is
position limited, obeying $P_s\vert\Psi\rangle=\vert\Psi\rangle$.

As was shown in \cite{ak-beet,Kempf2003}, it is always possible to find
incident wave functions $\psi(x)$ which obey the momentum bound $\pmax$
and which at any finite number of points in the slit interval take
arbitrarily prescribed amplitudes (we will reproduce this result as a
special case below). We will be able to arrange, therefore, that the wave
function $\psi(x)$ takes for example the alternating values $(-1)^n$ at an
arbitrarily large number of points in the slit interval - which enforces
superoscillations. These $\psi(x)$ will be differentiable and square
integrable. Then, if the particle passes the slit, only the
superoscillating stretch of the wave function emerges from the slit.
Renormalized, we denote it $\Psi(x)$. The Fourier transform $\ft{\Psi}(p)$
of $\Psi(x)$ will show the presence of small wavelengths, implying that
the particle emerges from the slit accelerated to a momentum beyond
$\pmax$.

\subsection{Template Functions}

As already mentioned, the results of \cite{ak-beet,Kempf2003} showed that
functions of fixed bandwidth can always be found which at arbitrarily but
finitely many points possess predetermined amplitudes. Therefore, the
width or narrowness of the slit does not limit how short the wavelength of
the superoscillations can be. As a consequence, there is no slit-dependent
limit to the amount of self-acceleration that can be achieved in this way.

This leads us to ask more generally whether the process of
self-acceleration can be designed virtually at will: is it always possible
to construct incident wave functions $\psi(x)$ of fixed momentum limit
$\pmax$ which on the slit interval $\slitinterval$ match any
arbitrarily-chosen template function, say $\Phi(x)$? This is of interest
because, if true, we can optimize the predictability of the
self-acceleration. To this end, we would choose the template function
$\Phi(x)$ to be a wave function with a fixed arbitrarily large momentum
expectation $\bar{p}$ whose momentum uncertainty $\Delta p$ is as small as
allowed by the uncertainty relation. If the incident superoscillatory wave
function matched this template function in the slit interval (up to
normalization), then the wave function would merge from the slit with the
chosen momentum expectation $\bar{p}$ and lowest possible momentum
uncertainty, $\Delta p$, for the given width of the slit. For later use,
let us calculate these ideal template functions $\Phi(x)$.

\subsubsection{Ideal Template Functions}
\label{idtem}

Our aim is to find `ideal' template functions $\Phi(x)$ defined on
the slit interval $\slitinterval$ which minimize the momentum
uncertainty $\variance{p}$, possess a predetermined momentum
expectation $\ip{\Phi}{\p\vert \Phi} = \bar{p}$ and are normalized
$\ip{\Phi}{\Phi} = 1$. To this end, we need to solve the
constrained variational problem with the functional
\begin{equation}
    \mathcal{L} = \langle \Phi\vert\p^2\vert\Phi\rangle +\mu_1
    \langle\Phi\vert\p\vert\Phi\rangle +\mu_2\langle\Phi\vert\Phi\rangle
\end{equation}
where $\mu_{1}$ and $\mu_{2}$ are Lagrange multipliers. Note that
minimizing $\variance{p}$ is equivalent to minimizing
$\ip{\Phi}{\p^{2}\vert\Phi}$ because $\expectation{p}$ is fixed. Hence the
Euler-Lagrange equation in position space is:
\begin{equation} \label{E:idealtemplateELE}
    -\hbar^{2}\Phi''(x) - i\hbar\mu_{1}\Phi'(x) + \mu_{2}\Phi(x) = 0
\end{equation}
Since any wave function that emerges from the slit vanishes at the slit
boundaries, we require $\Phi(\pm L/2)=0$. The solution is unique up to a
phase:
\begin{equation} \label{E:idealtemplate}
    \Phi(x) = \sqrt{\frac{2}{L}}~
    \cos{\frac{\pi}{L}x}~e^{ix\bar{p}/\hbar}
\end{equation}
Its uncertainties are $\Delta x= L ((\pi^2-6)/12 \pi^2)^{1/2}\approx 0.18
L$ and $\variance{p} = \frac{\pi\hbar}{L}$. We have $\Delta x \Delta p
\approx 0.57 \hbar$ which is a little larger than what the uncertainty
relation allows because our problem requires template functions $\Phi(x)$
to be zero outside the slit interval.

\subsubsection{Superoscillatory Wave Functions
Cannot Match Arbitrary Templates}

Let us now come back to the question whether it is generally possible to
find an incident wavefunction $\psi(x)$ which obeys a momentum bound
$\pmax$ while in the interval $\slitinterval$ agreeing completely with an
arbitrarily chosen template function $\Phi(x)$, such as the function
$\Phi(x)$ just calculated in Sec.\ref{idtem}. Strictly speaking, the
answer is no:

As is easily verified, all bandlimited functions are entire functions. In
particular, any momentum-limited incident wave function $\psi(x)$ is
entire and it is, therefore, everywhere differentiable. Now choose, for
example, a template function $\Phi(x)$ which is not differentiable at some
point in the interval $\slitinterval$. Thus, there cannot exist a
momentum-limited incident wave function which obeys $\psi(x)=\Phi(x)$ for
all $x\in \slitinterval$. Nevertheless, a slightly weaker proposition does
hold.

\subsubsection{Convergence Towards Arbitrary Template Functions}
\label{lemmasec}

Let $\Phi(x)$ be a continuous and square integrable template function. Let
us ask whether one can always find a sequence of wave functions
$\psi_{N}(x)$ of fixed momentum bound $\pmax$ which behave with more and
more precision like $\Phi(x)$ over the region of the slit. To be precise,
is it possible to find a sequence of momentum-limited incident
waves $\vert\psi_N\rangle$
whose emerging wave functions $\vert\Psi_N\rangle$ have asymptotically
vanishing $L^2$-distance $\normtwo{\kPsiN-\kPhi}$ to an arbitrary template
state $\vert\Phi\rangle$? This is indeed the case.

To see this, consider in the quantum mechanical Hilbert space of states
$\mathcal{H}$ with scalar product
\begin{equation}
    \ip{\xi_1}{\xi_2}:=\int_{-\infty}^\infty \xi_1^*(x)\xi_2(x)~dx
\end{equation}
and the following three subspaces:
\begin{align}
    \mathcal{H}_{s} &:= P_{s}~\mathcal{H} \\
        %\setbrackets{ ~\Psi\in\mathcal{H}~|~\supp \Psi(x) \subset\slitinterval~ } \\
    \mathcal{H}_{b} &:= P_{b}~\mathcal{H} \\
        %\setbrackets{ ~\Psi\in\mathcal{H}~|~\supp \\ft{\Psi}(p) \subset\pmaxinterval~ } \\
    \mathcal{H}_{sb} &:= P_{s}~P_{b}~\mathcal{H}
        %\setbrackets{ ~\Psi(x) = P_{L/2}~h(x)~|~h \in \mathcal{H}_{b}~}
\end{align}
That is, $\mathcal{H}_{s}$ is the subspace of states with
position-limitation to the slit, $\mathcal{H}_{b}$ is the subspace of
states with fixed momentum-limitation $\pmax$, and $\mathcal{H}_{sb}$ is
the subspace of states obtained after passing the momentum-limited wave
functions through the slit.
\begin{prop}
$\mathcal{H}_{sb}$ is dense in $\mathcal{H}_{s}$, i.e.:
\begin{equation}
    \forall \ \kPhi \in \mathcal{H}_{s}, \varepsilon > 0 \ \exists \ \kPsi \in
    \mathcal{H}_{sb} : \normtwo{\kPsi-\kPhi} < \varepsilon.
\end{equation}
\end{prop}
\begin{proof}
If $\kPhi=0$, then take $\kPsi=0$. For $\kPhi \ne 0$ we must show
that:
\begin{equation}
    \nexists \ \kPhi \in \mathcal{H}_{s}
    \backslash \{0\} : \ip{\Psi}{\Phi} = 0 \ \forall \ \kPsi \in \mathcal{H}_{sb}
\end{equation}
Since $\kPhi$ is position-limited, this is equivalent to showing
that:
\begin{equation}
    \nexists \ \kPhi \in \mathcal{H}_{s}
    \backslash \{0\} : \ip{\psi}{\Phi} = 0 \ \forall \ \kpsi \in \mathcal{H}_{b}
\end{equation}
Assume, for a contradiction, that:
\begin{equation}
    \exists \ \kPhi \in \mathcal{H}_{s} \backslash \{0\} :
    \ip{\psi}{\Phi} = 0 \ \forall \ \kpsi \in \mathcal{H}_{b}
\end{equation}
This implies that $\kPhi \perp\mathcal{H}_{b}$.  Thus, $\ft{\Phi}(p)=0$ on
$\pmaxinterval$. But, since $\ft{\Phi}(p)$ is entire and zero over a
finite interval, $\ft{\Phi}(p)=0$ everywhere on $\mathbb{R}$, i.e.,
$\vert\Phi\rangle=0$. This is a contradiction. Therefore,
$\mathcal{H}_{sb}$ is dense in $\mathcal{H}_{s}$.
\end{proof}
While this result proves the existence of bandlimited functions that are
arbitrarily close in the $L^2$ topology to any template function within
the window of the slit, the result does not provide explicit methods for
constructing such bandlimited functions.

\section{\bf{Constructive Method for General Linear Constraints}} \label{S:linearconstraints}

We now focus on practical methods for calculating superoscillatory wave
functions that approximate template functions in the slit interval. We
begin with the method for constructing superoscillatory functions
presented in \cite{Kempf2003}. This method allows one to
 specify that the to-be-found superoscillatory function takes arbitrarily
 chosen amplitudes $a_k$ at any finite number $N$
 of arbitrarily chosen points $x_k$:
\begin{equation} \label{E:pointconstraint}
    \psi^{(u)}(x_{k}) = a_{k}  \mbox{\qquad for~ } k=1,...,N
\end{equation}
The superscript ${}^{(u)}$ is to indicate that the function will generally
be unnormalized. Eq.\ref{E:pointconstraint} specifies a function which
possesses a superoscillating stretch. For example, we may choose the $x_k$
spaced closer than $h/\pmax$ and the amplitudes alternating, e.g.
$a_k=(-1)^k$. The normalized wave function $\psi(x) =
\psi^{(u)}(x)/\vert\vert\psi^{(u)}\vert\vert$ then possesses
superoscillations that are as rapid as those of $\psi^{(u)}$ but
 with a renormalized amplitude.  Thus, in
 order to obtain the $\psi(x)$ with the most
 pronounced superoscillations, i.e. the superoscillations
  of largest possible amplitude, one needs to find that function
$\psi^{(u)}(x)$ whose $L^2$ norm $\vert\vert \psi^{(u)}\vert\vert$ is
minimal. The method of \cite{Kempf2003}
 solves this optimization problem.

We now generalize the method of \cite{Kempf2003}. To this end, we begin by
rewriting the requirement that $\psi^{(u)}(x)$ be bandlimited by $\pmax$
and pass through the points $\{(x_{k},a_{k})\}_{k=1}^{N}$, namely
(\ref{E:pointconstraint}), in momentum space as
\begin{equation}
    \frac{1}{\sqrt{2\pi\hbar}} \pmaxint e^{i\frac{x_{k}}{\hbar}p}~\ft{\psi}^{(u)}(p)~
     ~ dp = a_{k}.
\end{equation}
Our aim is to obtain a method for constructing superoscillatory wave
functions which not only pass through predetermined points but which obey
also more generic types of constraints. To this end, let us allow
constraints on the function $\psi^{(u)}$ which are of the general linear
form:
\begin{equation} \label{E:constraints}
    a_{k} = \frac{1}{\sqrt{2\pi\hbar}} \pmaxint
    \chi^*_{k}(p) \ft{\psi^{(u)}}(p) \ dp \ \qquad \forall \ k \in \{1,\ldots,N\}
\end{equation}
Here, the $\chi_{k}$ are arbitrary linearly independent differentiable
functions. By choosing these, we will be able to prescribe for the
superoscillatory wave function not only amplitudes but also arbitrary
derivatives, integrals and any other linear constraint. In order to obtain
the most pronounced superoscillations in the normalized function $\psi$ we
minimize the norm of $\psi^{(u)}$, subject to the constraints in
Eq.\ref{E:constraints}. The to-be-optimized functional with Lagrange
multipliers $\lambda_k$ reads
\begin{equation}
    \mathcal{L} = \pmaxint {\psi^{(u)}}^*(p) \ft{\psi}^{(u)}(p) \
     dp - \sum_{k=1}^{N} \frac{\conj{\lambda_{k}}}{\sqrt{2\pi\hbar}}
     \pmaxint \chi^*_{k}(p) \ft{\psi}^{(u)}(p) \ dp + c.c.,
\end{equation}
leading to the Euler-Lagrange equation:
\begin{equation} \label{E:ftpsi}
    \ft{\psi}^{(u)}(p) = \frac{1}{\sqrt{2\pi\hbar}} \sum_{k=1}^{N}
    \lambda_{k} {\chi_{k}}(p)
\end{equation} Recall that $\ft{\psi}^{(u)}$
is zero outside the interval $\pmaxinterval$ by assumption. Thus, using
(\ref{E:ftpsi}) in (\ref{E:constraints}),
\begin{equation} \label{E:ak}
    a_{k} = \sum_{r=1}^{N} T_{kr} \lambda_{r},
\end{equation}
where the Hermitian matrix $T$ is defined by:
\begin{equation} \label{E:Tmatrix}
    T_{kr} := \frac{1}{2\pi\hbar} \pmaxint \chi^*_{k}(p)~{\chi_{r}}(p) ~dp
\end{equation}
As we will show below, $T$ is invertible. Thus, $\vec{\lambda} = T^{-1}
\vec{a}$, i.e.:
\begin{equation}
        \lambda_{k} = \sum_{r=1}^{N} T_{kr}^{-1} a_{r}
\end{equation}
Thus, using the Fourier transform of the constraint function
\begin{equation} \label{E:positionbasis}
    \chi_{k}(x) := \frac{1}{2\pi\hbar} \pmaxint
    {\chi_{k}}(p)~e^{\frac{ixp}{\hbar}}~dp
\end{equation}
we obtain from (\ref{E:ftpsi}) that the desired superoscillatory (still
unnormalized) incident wave function in position space is given by:
\begin{equation} \label{E:psi}
    \psi^{(u)}(x) = \frac{1}{\sqrt{2\pi\hbar}} \sum_{k=1}^{N} \lambda_{k} \chi_{k}(x)
\end{equation}

\subsection{Existence of the Solution}

It remains to be shown that $T$ is indeed invertible. To see this, let
$\vec{u}$ be an arbitrary vector. Then:
\begin{align}
    \vec{u}^{\dag} T \vec{u}
        &= \sum_{k,r=1}^{N} \conj{u_{k}} T_{kr} u_{r} \notag \\
        &= \frac{1}{2\pi\hbar} \pmaxint \sum_{k,r=1}^{N}
        \conj{u_{k}} \conj{\chi_{k}}(p) \chi_{r}(p) u_{r}
        \ dp \notag \\
        &= \frac{1}{2\pi\hbar} \pmaxint \abs{ \sum_{m=1}^{N}
        u_{m} \chi_{m}(p) }^{2}  \ dp \notag
\end{align}
Since the $\chi_{k}$ are linearly independent the integrand is positive.
Therefore, $T$ is positive definite and hence invertible.

\section{{\bf The Cost of Superoscillations}}

As was shown in \cite{Kempf2003}, one cost of superoscillations is
that requiring more or faster superoscillations makes the matrix
$T$ increasingly difficult to invert numerically, as its smallest
and largest eigenvalues differ by growing orders of magnitude. The
condition number was found to increase exponentially with the
number of superoscillations.

We here only remark that, in the sense of computational complexity, this
makes it computationally hard to calculate superoscillations.
Interestingly, this also means that any quantum effect that naturally
produces functions with arbitrarily large superoscillatory stretches
constitutes an example of an exponential speed-up in the sense of quantum
computing. Physical occurrences of superoscillations, e.g. in the context
of evanescent waves, have been discussed e.g. in
\cite{Berry1994,Berry1995}. Also, for example, (rather speculatively) the
possibility of an unbounded production of superoscillations has been
discussed in the context of the transplanckian problem of black holes in
\cite{Reznik1997,Rosu1997}.

Here, we will focus on a more immediate cost of superoscillations, namely
the need for an increasingly large dynamical range: a function's
superoscillations are generally of low amplitude when compared to the
function's amplitudes to the left and right of its superoscillatory
stretch. To be precise, it was shown in \cite{Kempf2003} that the $L^2$
norm of the function increases polynomially with the frequency of the
prescribed superoscillations, for fixed prescribed superoscillating
amplitudes. In particular, it was also shown that the norm increases
exponentially with the number of imposed superoscillations.
Correspondingly, in normalized wave functions the amplitudes of
superoscillations decrease exponentially with the number of
superoscillations. (Of course, if the superoscillating stretch of the
particle's wave function happens to pass through the slit then its wave
function, however small, is re-normalized whereby the superoscillating
amplitudes will be restored to the amplitudes of the template function.)

By making use of the special properties of prolate functions these scaling
results were derived for the type of superoscillations produced with
method of \cite{Kempf2003}. In the following two subsections we will show
more directly the underlying reason for this exponential behavior of the
norm of superoscillatory functions. Our argument will apply more generally
to all superoscillatory functions that arise from linear constraints.

\subsection{Derivatives and Norms}

If a function $\psi^{(u)}(x)$ is bandlimited one would expect that there
is a bound on its derivatives.  Applying the Cauchy-Schwarz inequality,
consider
\begin{align}
    \abs{\frac{d^{n}}{dx^{n}}\psi^{(u)}(x)}^{2}
        &= \frac{1}{2\pi\hbar} \abs{ \pmaxint \ft{\psi^{(u)}}(p)
        \roundbrackets{\frac{ip}{\hbar}}^{n} e^{\frac{ipx}{\hbar}} \ dp }^{2}
         \notag \\
        &\leq \frac{1}{2\pi\hbar} \roundbrackets{
        \pmaxint \abs{\ft{\psi^{(u)}}(p) \roundbrackets{
        \frac{ip}{\hbar}}^{n} }^{2} \ dp } \roundbrackets{ \pmaxint
        1 \ dp } \notag \\
        &\leq \frac{1}{2\pi\hbar}
         \roundbrackets{\frac{\pmax}{\hbar}}^{2n} 2\pmax \normtwo{\psi^{(u)}}^{2}. \notag
\end{align}
Thus,
\begin{equation}
    \abs{\frac{d^{n}}{dx^{n}}\psi^{(u)}(x)} \leq
    \roundbrackets{\frac{\pmax}{\hbar}}^{n} \sqrt{\frac{\pmax}{\pi\hbar}}
    \normtwo{\psi^{(u)}}.
\end{equation}
Thus arbitrarily large derivatives, as they can be produced with
superoscillations, are consistent with a finite fixed bandwidth but we see
that the cost must be an increase in the norm of the function
$\normtwo{\psi^{(u)}}$.

\subsection{The Norm of Superoscillating Functions}

A precise expression for the norm $\normtwo{\psi^{(u)}}$ of the
superoscillatory functions obtained by our method can be derived:
\begin{align}
    \normtwo{\psi^{(u)}}^{2}
        &= \frac{1}{2\pi\hbar} \pmaxint \abs{ \sum_{k=1}^{N}
        \lambda_{k} \chi_{k}(p) }^{2} \ dp \notag \\
        &= \frac{1}{2\pi\hbar} \pmaxint \sum_{k,r=1}^{N}
        \conj{\lambda_{k}} \conj{\chi_{k}}(p) \chi_{r}(p)
        \lambda_{r} \ dp \notag \\
        &= \vec{\lambda}^{\dag} T \vec{\lambda}\label{abc}
\end{align}
Hence:
\begin{equation} \label{E:norm}
    \normtwo{\psi^{(u)}}^{2} = \vec{\lambda}^{\dag} \vec{a}
    = \vec{a}^{\dag} \vec{\lambda} = \vec{a}^{\dag} T^{-1} \vec{a}
\end{equation}
Note that $T^{-1}$ is a positive self-adjoint matrix. We now see that for
given constraint functions, $\chi_k$, the most norm-expensive
superoscillatory functions are obtained if we choose the constraint
parameters $a_k$ such that $\vec{a}$ is the eigenvector of $T^{-1}$ with
largest eigenvalue. We will arrive at those extreme superoscillations also
from independent momentum space considerations in Sec.\ref{sectm}.
\label{prom}

\subsection{Adding Successive Constraints}
\label{succ}

Consider a set of constraints, described by a set of functions
$\{\chi_{k}\}_{k=1}^{N}$ and parameters $\{a_{k}\}_{k=1}^{N}$ and
suppose that, using our method, the momentum-limited wave function
which obeys all those constraints and is of minimum norm has been
calculated. Let us ask how the norm of the solution to this
problem changes if we add one additional constraint
\begin{equation}
    a_{N+1} = \frac{1}{\sqrt{2\pi\hbar}} \pmaxint
    \chi_{N+1}(p) \ft{\psi}(p) \ dp
\end{equation}
where $\chi_{N+1}$ and $a_{N+1}$ are chosen arbitrarily. Let us denote the
solution to the initial problem of $N$ constraints by $\psi_{N}$ and let
us define:
\begin{equation} c:=
\frac{1}{\sqrt{2\pi\hbar}} \pmaxint
    \chi_{N+1}(p) \ft{\psi_N}(p) \ dp
\end{equation}
Clearly, if we choose the $(N+1)$st constraint with $a_{N+1}:=c$
then $\psi_{N}$ is also the function of minimum norm obeying the
$N+1$ constraints, i.e. $\psi_{N+1}=\psi_{N}$, just as if we had
not added a new constraint, or as if we had set the $(N+1)$st
Lagrange multiplier to zero: $\lambda_{N+1}=0$.

Now, let us allow the constraint parameter $a_{N+1}$ to vary away from
$c$. Correspondingly, our method will yield a family of functions,
$\psi_{N+1}~~(\neq\psi_N)$, parametrized by $a_{N+1}$. We observe from
(\ref{E:norm}) (letting the sum run to $N+1$) that the norm squared of
these functions is a quadratic (and of course positive) polynomial in
$a_m$. Note that its minimum occurs if we choose the $a_{N+1}$-value
\begin{equation}
  c = a_{N+1} = -\frac{1}{T_{(N+1),(N+1)}^{-1}} \sum_{ r \ne (N+1) }
    T_{(N+1),r}^{-1} a_{r}
\end{equation}
because then $\sum_{s\neq (N+1)}T^{-1}_{(N+1),s}a_s=0$. Using (\ref{E:ak})
we see that this choice of $a_{N+1}$ leads to the vanishing of the
Lagrange multiplier $\lambda_{N+1}=0$, which is what we expected for if we
add a new constraint that is already satisfied, $\psi$ will not change.

Crucially, we now see that as we tune $a_{N+1}$ away from $c$, say in
order to enforce an additional superoscillation twist, the squared norm of
the solution increases quadratically. Therefore, if we keep adding new
generic constraints, say in order to implement more and more
superoscillations, this will generally increase the norm of the solution
by a factor in each step. Thus, the norm of the solution will generically
scale exponentially with an increase in the number of constraints $N$.

This finding widely generalizes the result of \cite{Kempf2003} which
applied only to constraints of the special form (\ref{E:pointconstraint})
and among them only to those with equidistant spacings of the $x_k$.

\section{\bf{Applications to an `Ideal' Template Function}}

In Sec.\ref{idtem}, we asked how the wave function $\Psi(x)$ that emerges
from the slit would have to look in order to describe a particle with an
arbitrarily high predetermined momentum expectation value $\bar{p}$ and a
momentum uncertainty $\Delta p$ which is as small as is allowed by the
uncertainty relation. This `ideal' template function was given in
(\ref{E:idealtemplate}).

\begin{figure}[h]
\centerline{\includegraphics[width=\figwid,angle=-90]{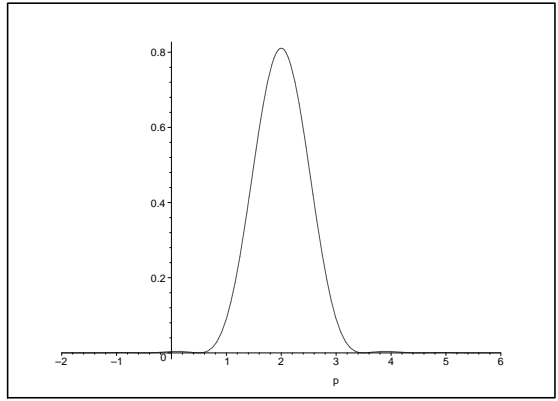}}
    \caption{\label{F:Phi_abs.eps} The `ideal' template's momentum wave function
    $\abs{\Phi(p)}^2$. Notice that it is centred well outside the
    original bandwidth $\pmax=1$.}
\end{figure}

Let us consider the concrete example, $\hbar=1$, $L=2\pi$, $\pmax=1$, and
$\bar{p}=2$.
%In this case, the shortest wavelength occurring in the Fourier transform
%is $\lmin=2\pi$ which equals the width of the slit.
If the emerging wave function $\Psi$ can be arranged to be equal
or close to this template function $\Phi$, this clearly exhibits
the phenomenon of self-acceleration because the emerging momentum
wave function would be peaked at $p\approx \bar{p}=2$, i.e. well
outside the original bandwidth of $\pmax=1$, see
Fig.\ref{F:Phi_abs.eps}.

$$$$
We had shown that exact matching, $\Psi(x)=\Phi(x)$, is generally not
possible, but we also saw that there always exists a sequence of incident
waves $\psi_N$ so that for the emerging waves $\Psi_N$ we have $\Psi_N
\rightarrow \Phi$ in the $L^2$ topology, which is here the only physically
relevant topology. Thus, there are superoscillatory incident wave
functions which achieve the prescribed self-acceleration properties to
arbitrary precision. For illustration, let us explicitly calculate such
superoscillatory incident wave functions.

\subsection{Method of Matching Amplitudes}

\label{exam:points} Let us begin by applying the method presented
in \cite{Kempf2003}, which is a special case of our method of
general linear constraints. In this special case, we require the
momentum-limited incident wave $\psi^{(u)}(x)$ to exactly match
the amplitude of the ideal template function at several points
$x_k$ of the slit interval $\slitinterval$. The constraints in the
variational problem are then given by the linearly independent
constraint functions $ \chi_{k}(p):=e^{-\frac{ipx_{k}}{\hbar}}$
and constraint parameters $ a_{k}:=\Phi(x_{k})$. Thus,
\begin{align}
    T_{kr}
        &= \frac{1}{2\pi\hbar} \pmaxint e^{\frac{ip}{\hbar}(x_{k}-x_{r})} \ dp \notag \\
        &= \frac{\pmax}{\pi\hbar} \ \sinc{\frac{\pmax}{\hbar}[x_{k}-x_{r}]}
\end{align}
which leads to the solution:
\begin{equation} \label{E:psipoints}
    \psi^{(u)}(x) = \frac{1}{\sqrt{2\pi\hbar}} \sum_{k=1}^{N} \lambda_{k} \chi_{k}(x),
\end{equation}
where $\vec{\lambda}=T^{-1}\vec{a}$ and where
\begin{align}
    \chi_{k}(x)
        &= \frac{1}{\sqrt{2\pi\hbar}} \pmaxint e^{\frac{ip}{\hbar}(x-x_{k})} \ dp \notag \\
        &= \pmax \sqrt{\frac{2}{\pi\hbar}} \ \sinc{\frac{\pmax}{\hbar}[x-x_{k}]}.
\end{align}
We observe that the wave function $\psi^{(u)}(x)$ is a linear
combination of sinc functions centred at the $x_{k}$ and we note
that $\psi(x)$ is square integrable, since the sinc functions are.
In general, $T$ ill-conditioned, i.e. care must be taken to invert
it with enough numerical precision so as to satisfy the
constraints with sufficient accuracy.

\begin{figure}[h]
    \centerline{\includegraphics[width=\figwid,angle=-90]{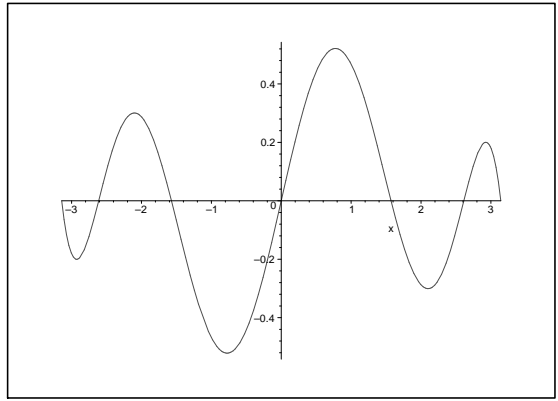}}
    \caption{\label{F:points_psi_i}$\Im{\psi^{(u)}(x)}$ over the slit in
    the
example of Sec.\ref{exam:points}.  The wavelength is about $0.5\lmin$.}
\end{figure}

\begin{figure}[h]
\centerline{\includegraphics[width=\figwid,angle=-90]{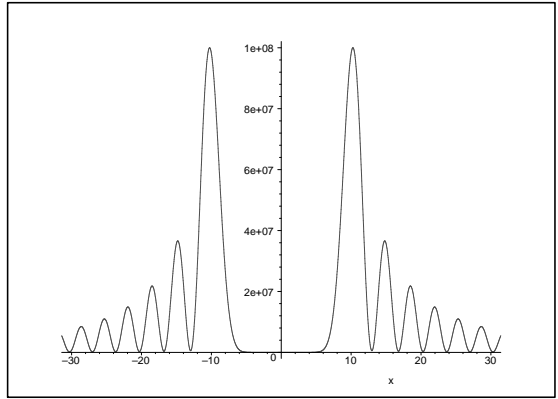}}
    \caption{\label{F:points_psi_abs}$\abs{\psi^{(u)}(x)}^{2}$ over the slit and
surrounding regions in the example of Sec.\ref{exam:points}.}
\end{figure}

We used routines in Maple which calculate
$\vec{\lambda}=T^{-1}\vec{a}$ by solving $T\vec{\lambda}=\vec{a}$
using Gaussian elimination. Concretely, we required
$\psi^{(u)}(x)$ to match the ideal template function $\Phi(x)$
with $\bar{p}=2$ at $N=9$ equidistantly-spaced points $x_k$ from
slit boundary to slit boundary. For example,
Fig.\ref{F:points_psi_i} shows the imaginary part of the
superoscillatory function $\psi^{(u)}$ over the slit interval.
Fig.\ref{F:points_psi_abs} shows a zoomed-out view of
$\abs{\psi^{(u)}(x)}^{2}$, displaying the typical big amplitudes
to the left and right of the superoscillating stretch.


The momentum expectation value for the ideal template function
that we chose is $\bar{p}=2\pmax$. Numerically, we found that the
strictly momentum limited incident wave function $\psi(x)$ for
$N=9$ yields an emerging wave function $\Psi(x)$ whose momentum
expectation value is $\bar{p}\approx 1.92\pmax$. Clearly, the
momentum of particles which pass through the slit essentially
doubles by self-acceleration, as intended. The momentum
uncertainty of the emerging wave function is $\Delta p \approx
1.42 \pmax$.

Recall that for this slit size the momentum uncertainty could be
significantly smaller, namely $\Delta p=1/2$, as is precisely realized in
the ideal template function. By increasing $N$, we can achieve that the
incident wave function $\psi^{(u)}(x)$ better matches the template,
leading to a lowering of $\Delta p$ towards that limiting value. For
example, for $N=15$ we find $\bar{p}\approx 1.99947 \pmax$ and $\Delta p
\approx 0.50025 \pmax$. For significantly larger $N$ the exponential
computational expense sets in. Our generalized method for linear
constraints allows us to use other linear constraints which we found to be
numerically more efficient in the sense of allowing us to reach larger
values of $N$. We will discuss the use of these alternative constraints in
Sec.\ref{exam:derivs}.

\begin{figure}[h]
    \centerline{\includegraphics[width=\figwid,angle=-90]{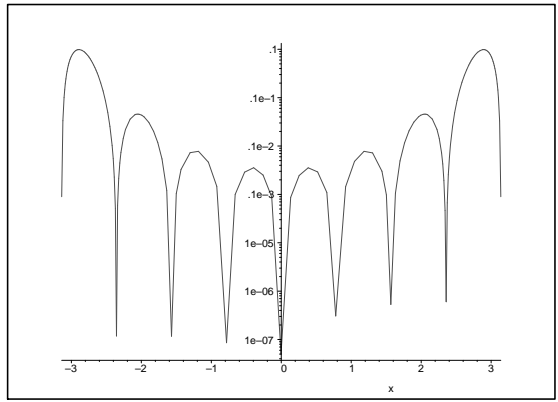}}
    \caption{\label{F:points_diff}$\abs{\Psi(x)-\Phi(x)}^{2}$ over the slit
in the example of Sec.\ref{exam:points}.}
\end{figure}
Fig.\ref{F:points_diff} shows the accuracy with which the $\psi^{(u)}(x)$
obtained by matching $N=9$ amplitudes of $\psi^{(u)}(x)$ to those of
$\Phi(x)$ agrees with the ideal template function $\Phi(x)$ for arbitrary
$x$ in the slit interval. The behavior is similar for all values of $N$
(that we tested). For general $N$, there are $N-1$ peaks and the height of
the highest peak decreases with $N$.

\subsection{Method of Matching Derivatives}
\label{exam:derivs}

In order to illustrate the generality of our new method of
Sec.\ref{S:linearconstraints}, let us now construct superoscillatory wave
functions by requiring that the wave function matches value and
derivatives of the template at one point, instead of requiring, as we did
in Sec.\ref{exam:points}, that the wave function matches only the value of
the template function at several points.

Concretely, let us require that the value and first $N-1$ derivatives of
the to-be-found wave function $\psi^{(u)}$ agree at $x=0$ with those of
the `ideal' template function $\Phi$ of above. In the equation for general
linear constraints, (\ref{E:constraints}), we obtain a constraint on the
$(k-1)$st derivative by choosing for the constraint function:
\begin{equation}
    \chi_{k}(p)=\roundbrackets{-\frac{ip}{\hbar}}^{k-1}
\end{equation}
Matching the derivatives to those of the template is to choose the
constraint parameters to be $a_{k}:=\Phi^{(k-1)}(0)$, where $\Phi^{(k-1)}$
denotes the $(k-1)$st derivative. Since the $\chi_{k}$ are linearly
independent,
\begin{equation}
    T_{kr} = \frac{1}{2\pi\hbar} \pmaxint
    \roundbrackets{\frac{ip}{\hbar}}^{k-1}
    \roundbrackets{-\frac{ip}{\hbar}}^{r-1} \ dp
\end{equation}
is invertible, yielding the solution
\begin{equation}
    \psi^{(u)}(x) = \frac{1}{\sqrt{2\pi\hbar}} \ \sum_{k=1}^{N} \lambda_{k} \chi_{k}(x),
\end{equation}
where
\begin{equation}
    \chi_{k}(x) = \frac{1}{\sqrt{2\pi\hbar}} \pmaxint \roundbrackets{-\frac{ip}{\hbar}}^{k-1} e^{\frac{ipx}{\hbar}} \ dp.
\end{equation}
Note that $\psi(x)$ is a linear combination of derivatives of sinc
functions, each of which is bandlimited. In this case, $T$ is simpler to
invert and we can go, for example, to the case $N=23$ before the
exponential computational expense sets in. In this case, for large $N$ the
coefficients $\lambda_{k}$ quickly grow large and hence the subtle
cancellations in the Fourier transform require fast increasing numerical
precision.

\begin{figure}[h]
    \centerline{\includegraphics[width=\figwid,angle=-90]{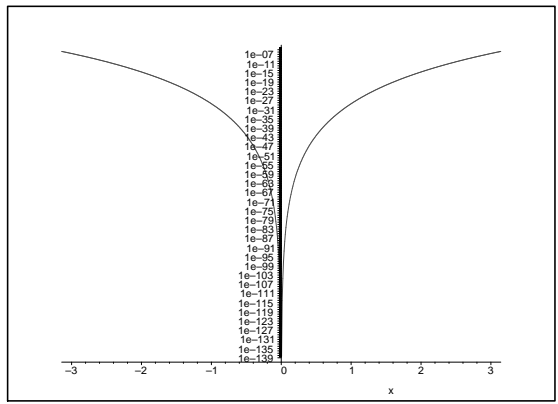}}
    \caption{\label{F:derivs_diff}$\abs{\Psi(x)-\Phi(x)}^{2}$ over the slit
in the example of Sec.\ref{exam:derivs}.}
\end{figure}

We considered the example where the value of the function and its first 22
derivatives is required to match those of the ideal template function at
$x=0$. We found numerically that the momentum-limited superoscillating
function, after passing through the slit, then exhibits a momentum
expectation value of $\bar{p}\approx 2.0002 \pmax$ and momentum
uncertainty $\Delta p\approx 0.50049$. Thus we reach the targeted
momentum-doubling self-acceleration, with a momentum uncertainty which is
only marginally above the uncertainty relation limit $\Delta p=1/2$ for
this slit size. Fig.\ref{F:derivs_diff} displays the accuracy
$\abs{\Psi(x)-\Phi(x)}^{2}$.

\section{\bf{A Momentum Space Method}}
\label{sectm}
In position space, superoscillatory wave functions $\psi(x)$
generally possess a characteristic shape: rapid but small oscillations in
the superoscillating stretch and a few large long-wavelength amplitudes
shortly before and after. Do these states also possess a characteristic
shape in momentum space?

\begin{figure}[h]
\centerline{\includegraphics[width=\figwid,angle=-90]{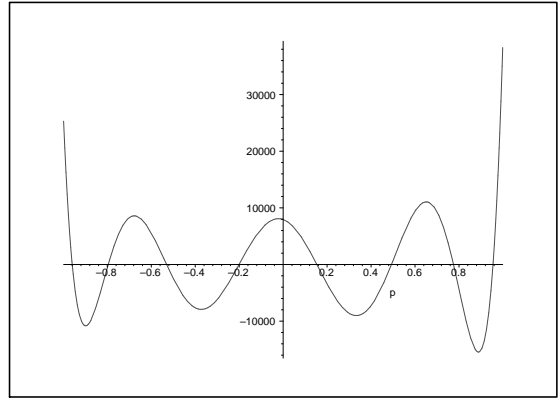}
}
    \caption{\label{F:points_psiF_r}\label{F:points_psi_four_r}
    $\Re{\ft{\psi}(p)}$ of the incident superoscillatory wave in
    the example in Sec.\ref{exam:points}.}
\end{figure}

Let us consider, for example, the superoscillations obtained by
prescribing oscillating amplitude values $a_k$ at close-by points $x_k$.
We found that, in momentum space, such a state is a linear combination of
plane waves $\exp(-ix_kp)$:
\begin{equation}
    \ft{\psi}(p) :=
    \begin{cases}
        \frac{1}{\sqrt{2\pi\hbar}} \sum_{r=1}^{N}
        \lambda_{r} e^{ -i \frac{x_{r}}{\hbar} p },
        \mbox{\quad if}~~\abs{p} \leq \pmax \\
        0, ~~\mbox{\qquad if}~~\abs{p} > \pmax
    \end{cases}
\end{equation}
It appears, see e.g. Fig.\ref{F:points_psi_four_r}, that these
$\psi(p)$ generally possess small amplitudes in most of the
momentum interval $\pmaxinterval$, except for the near the
boundaries $\pm\pmax$. We calculated the Fourier transforms of a
number of superoscillatory wave functions and observed this as a
general feature.

Thus, in momentum space, these superoscillations appear to be a linear
combination of plane waves whose interference is close to being as strong
as it can be, with the effect that the resulting function is of minimized
norm.

If this assumption is correct, we should be able to derive
superoscillatory wave functions by calculating that linear combination of
plane waves in momentum space whose norm is minimal. To this end, let
$\{x_{r}\}_{r=1}^{N}$ be points in $\slitinterval$. Our aim is to find a
coefficient vector $\{q_{r}\}_{r=1}^{N}$ of fixed length, say
$\normtwo{\vec{q}} = 1$ such that
\begin{equation}
    \ft{\psi}(p) :=
    \begin{cases}
        \frac{1}{\sqrt{2\pi\hbar}} \sum_{r=1}^{N}
        q_{r} e^{ -i \frac{x_{r}}{\hbar} p },
        \mbox{\quad if}~~\abs{p} \leq \pmax \\
        0, ~~\mbox{\qquad if}~~\abs{p} > \pmax
    \end{cases}
\end{equation}
is of minimum norm. The constrained optimization problem with Lagrange
multiplier $\nu$
\begin{equation}
    \mathcal{L} = \normtwo{\ft{\psi}}^{2} + \nu ( \normtwo{\vec{q}}^{2} - 1 )
\end{equation}
leads to:
\begin{equation}
        T \vec{q} = \nu \vec{q} \\
\end{equation}
Thus, the coefficient vector $\vec{q}$ which solves this optimization
problem must be eigenvector to $T$. From Eq.\ref{abc}, we obtain the
general expression for the norm: $\vert\vert\psi\vert\vert =
\vec{q}^\dagger T\vec{q}$. Thus, $\vec{q}$ must be that eigenvector of $T$
with the smallest eigenvalue.

Indeed, the position wave function determined by these coefficients
$\vec{q}$ is superoscillatory: already in Sec.\ref{prom}, we encountered
the wave functions whose coefficient vectors $\lambda$ are the
eigenvectors of $T^{-1}$ of largest eigenvalue. There, we found that these
are the superoscillatory wave functions which for a given set of
constraint points $\{x_k\}$ are most norm-expensive and which, therefore,
possesses the most pronounced superoscillations.

\section{{\bf Open Problems}}

We know from Sec.\ref{lemmasec} that it is always possible to find
incident wave functions of fixed momentum bound that in the slit
interval are matching any given template function arbitrarily
closely in the $L^2$ norm topology. Thus, for all practical
purposes, the self-acceleration phenomenon can be tailored at
will. Our method of general linear constraints can be used to
explicitly construct a sequence of momentum-limited
superoscillatory wave functions $\psi_N(x)$ which more and more
closely match any given template function $\Phi$. The $\psi_N(x)$
approach $\Phi(x)$ in the slit interval in the sense that they
obey more and more linear constraints that tie $\psi_{N}(x)$ to
$\Phi(x)$.

In Sec.\ref{exam:points} and Sec.\ref{exam:derivs} we showed that a close
approach to a fixed template function can be done numerically efficiently.
Clearly, intuition and the easily achieved numerical accuracy lead us to
conjecture that our methods for producing superoscillations, as used in
Secs.\ref{exam:points},\ref{exam:derivs}, do indeed always lead to
convergence in the $L^2$-topology towards the template function. So far,
however, we have no proof that our particular method for producing
superoscillatory wave functions from linear constraints does indeed
realize the $L^2$ convergence to generic template functions.

\subsection{Quadratic constraints}
Let us ask, therefore, if there is a choice of linear constraints that
directly targets the area under the functions and that thereby directly
guarantees convergence in the $L^2$ sense.

One may try, for example, constraints which require that the functions
$\psi^{(u)}$ and $\Phi$ enclose equal areas on certain subintervals of the
slit. This can be put into the form of a linear constraint: Let
$\{x_{k}\}_{k=1}^{N+1}$ be equidistantly-spaced points in $\slitinterval$.
We require the linear constraints of (\ref{E:constraints}) with the
constraint functions
\begin{equation}
    \chi_{k}(p) = \int_{x_{k}}^{x_{k+1}} e^{-\frac{ipx}{\hbar}} \ dx.
\end{equation}
and the constraint parameters:
\begin{equation}
    a_{k} = \int_{x_{k}}^{x_{k+1}} \Phi(x) \ dx
\end{equation}
While this can easily be carried out, these constraints are not
directly guaranteeing $L^2$ convergence towards the template
function by refining the partition of the slit interval into
increasingly smaller subintervals: in principle, even functions
that enclose equal areas on a very small interval may have very
different amplitudes. Let us, therefore, consider to impose
constraints which require the area of the function $\vert
\Psi(x)-\Phi(x)\vert^2$ on small subintervals to be small. It is
clear that to this end it would be necessary to implement also
constraints that are quadratic in the field $\Psi(x)$. We will
here not pursue this strategy to the end. As a preliminary step,
however, let us generalize our method for constructing
superoscillatory functions to include quadratic constraints.

To this end, we formulate the variational problem of finding the
function $\psi^{(u)}$ of smallest norm and with momentum cutoff
$\pmax$ which satisfies $N$ linear and $M$ quadratic constraints
that tie it to a template function $\Phi$:

\begin{eqnarray}
    a_{k} & = & \frac{1}{\sqrt{2\pi\hbar}} \pmaxint
    {\psi^{(u)}}(p) ~ \chi_{k}^{*}(p)
     \ dp \qquad \mbox{ for } k=1,...,N\label{ay}\\ \nonumber \\
    b_{k} & = & \frac{1}{\sqrt{2\pi\hbar}} \pmaxint
    \conj{{\psi^{(u)}}}(p) ~ \ft{\psi}^{(u)}(p) ~ \conj{\Xi_{k}}(p) \ dp
    \qquad \mbox{ for } k=1,...,M\label{az}
\end{eqnarray}
The to-be-optimized functional with Lagrange multipliers
$\{\lambda_{k}\}_{k=1}^{N}$ and $\{\mu_{k}\}_{k=1}^{M}$, reads
\begin{eqnarray}
    \mathcal{L} & = & \pmaxint {\psi^{(u)}}^*(p) ~ \ft{\psi}^{(u)}(p) \ dp
    - \sum_{k=1}^{N} \frac{{\conj{\lambda_{k}}}}{\sqrt{2\pi\hbar}}
    \pmaxint {\psi^{(u)}}(p) ~ \conj{\chi_{k}}(p)  \ dp \nonumber   \\ \nonumber \\
    &  & + \sum_{k=1}^{N} \frac{{\conj{\mu_{k}}}}{\sqrt{2\pi\hbar}}
    \pmaxint {\psi^{(u)}}^*(p) ~ \ft{\psi}^{(u)}(p)~\conj{\Xi_{k}}(p) \ dp + c.c.
\end{eqnarray}
and the Euler-Lagrange equation reduces to
\begin{equation}
    \ft{\psi}^{(u)}(p) = \frac{ \frac{1}{ \sqrt{2\pi\hbar} } ~ \sum_{k=1}^{N} \lambda_{k}
     {\chi_{k}}(p) }{ 1 + \frac{1}{ \sqrt{2\pi\hbar} } ~ \sum_{k=1}^{N}
     \mu_{k} {\Xi_{k}}(p) }.\label{ax}
\end{equation}

Although this may be difficult in practice, in principle, the substitution
of (\ref{ax}) into (\ref{ay}) and (\ref{az}) yields sufficient equations
to solve for the $\{\lambda_{k}\}_{k=1}^{N}$ and $\{\mu_{k}\}_{k=1}^{M}$
in terms of the $\{a_{k}\}_{k=1}^{N}$ and $\{b_{k}\}_{k=1}^{M}$ and this
yields the solution $\psi^{(u)}$.

\subsection{{\bf A Conjecture}}

\label{ac}

Consider the case of a differentiable template function $\Phi$ whose
derivative is bounded: $\abs{\Phi'(x)} \leq K \ \forall \ x \in
\slitinterval$, for some finite $K$. Assume that $\psi^{(u)}_N$ is a
sequence of incident wave functions, calculated through the method of
Sec.\ref{exam:points} with the amplitudes of $\psi^{(u)}_N(x)$ and
$\Phi(x)$ matched at $N$ equidistantly-spaced points $x_k$. We conjecture
then that the supremum $\abs{\psi'_{N}(x)}$ for all $x$ and all $N$ is
finite as well:
\begin{equation}
\abs{\psi_{N}'(x)} \leq M \ \quad \forall \ x \in \slitinterval,
\end{equation}
for some finite $M$. This is plausible because, else, $\abs{\psi'_{N}(x)}$
would have to develop arbitrarily sharp spikes away from the template
function in between some two points where its amplitudes are matched to
those of the template function. From Sec.\ref{succ}, however, we expect
large oscillations in the superoscillating stretch to be norm-expensive
and therefore be prevented from occurring, given that the $\psi^{(u)}_N$
are optimized to possess minimum norm for a given set of constraints.

\subsection{{\bf Convergence}}

\begin{prop} Assume that the conjecture of Sec.\ref{ac} holds true. Then,
$\{\psi_{N}(x)\}_N$ converges uniformly and in the $L^{2}$ topology over
the interval $\slitinterval$ to $\Phi(x)$ for $N\rightarrow\infty$.
\end{prop}
\begin{proof}
Partition the slit into $(N-1)$ equal-length intervals with the $N$
endpoints $x_k^{(N)} := -\frac{L}{2}+(k-1)\frac{L}{N-1}$. Define
$\{\alpha_{N}(x)\}_{N=2}^{\infty}$ by
\begin{equation}
    \alpha_{N}(x) := \max \{ ~x_{k}^{(N)}~ | ~k \in \{1,\ldots,N~ \},
    ~x_{k}^{(N)} \leq x~ \}.
\end{equation}
That is, $\alpha_{N}(x)$ is the closest point in the partition from the
left to $x$. Then,
\begin{align}
    \abs{\psi_{N}(x)-\Phi(x)}
        &\leq \abs{\psi_{N}(x)-\Phi(\alpha_{N}(x))} +
        \abs{\Phi(\alpha_{N}(x))-\Phi(x)} \notag \\
        &= \abs{\psi_{N}(x)-\psi_{N}(\alpha_{N}(x))} +
         \abs{\Phi(\alpha_{N}(x))-\Phi(x)} \notag \\
        &\leq M\abs{x-\alpha_{N}(x)} + K\abs{\alpha_{N}(x)-x} \notag \\
        &\leq \frac{(M+K)L}{N-1} \label{E:psigpointbound}
\end{align}
where we applied the triangle inequality and\texttt{} the mean value theorem. We
therefore have uniform and $L^2$-convergence.
\end{proof}

\section{Summary}

We started with the method for calculating superoscillatory wave functions
introduced in \cite{Kempf2003} and applied it to concrete examples. We
then generalized this method so that it now allows us to construct
superoscillatory low-momentum wave functions with a wide range of
predetermined properties. Namely, we can impose any arbitrary finite
number of linear constraints. We calculated concrete examples.

Further, we addressed the question whether superoscillatory functions can
be made to match any arbitrary continuous function on a finite interval.
This would correspond to imposing an infinite number of constraints.
Generally, the answer is no. However, we were able to prove that there
always exists a sequence of superoscillatory wave functions which
converges in the physically relevant $L^2$ topology towards any continuous
template function over an arbitrarily large chosen interval.

This is of interest for example in the case of the single slit: we proved
that the wave function of an incident low-momentum particle can be chosen
to arbitrary precision such that, if the particle passes through the slit,
it will emerge with a predetermined arbitrarily large momentum expectation
and with a momentum uncertainty that is as small as permitted by the width
of the slit.

Our method for constructing superoscillating wave functions allows us to
construct superoscillatory wave functions which match any \it finite \rm
number $N$ of properties of a given template function (such as the
template function's amplitudes or derivatives at specified points). This
leads to the question if by letting the number of constraints, $N$, go to
infinity we can obtain one of those sequences of superoscillatory wave
functions which converge towards the template function in the $L^2$
topology.

We proved that such sequences exist but we have not proved that our
particular method produces such sequences. The numerical evidence
certainly suggests that this is the case. In fact, we found rather fast
numerical convergence.

Nevertheless, it would be highly desirable to be able to prove that a
given method for producing superoscillations can be used to calculate a
sequence of superoscillatory functions that converges in the $L^2$
topology towards any given template function on an interval. An
investigation based on Weierstrass' approximation theorem is in progress,
\cite{kempf_bodmann}.

Lastly, we found a method for identifying a class of superoscillatory
functions by looking at their behavior in momentum space: superoscillatory
functions can be viewed as functions which in momentum space are a linear
combination of plane waves with coefficients such that their interference
is maximal, i.e. such that their norm is minimal.

$$$$

\noindent \bf Acknowledgement: \rm This work has been supported by NSERC,
CFI, OIT, PREA and the Canada Research Chairs Program.

\end{document}